\pdfoutput=1

\documentclass[3p,times,twocolumn]{elsarticle}

\usepackage{ecrc}

\volume{00}

\firstpage{1}

\journalname{Nuclear Physics B Proceedings Supplement}

\runauth{A. Schukraft for the IceCube Collaboration}

\jid{nuphbp}

\jnltitlelogo{Nuclear Physics B Proceedings Supplement}

\usepackage{amssymb}

\usepackage[figuresright]{rotating}

\begin{document}

\begin{frontmatter}



\dochead{}

\title{A view of prompt atmospheric neutrinos with IceCube}


\author{Anne Schukraft for the IceCube Collaboration}

\address{III. Physikalisches Institut, RWTH Aachen, D-52056 Aachen}

\begin{abstract}
Atmospheric neutrinos are produced in air showers, when cosmic-ray primaries hit the Earth's atmosphere and interact hadronically. The conventional neutrino flux, which dominates the neutrino data measured in the GeV to TeV range by neutrino telescopes, is produced by the decay of charged pions and kaons. Prompt atmospheric neutrinos are produced by the decay of heavier mesons typically containing a charm quark. Their production is strongly suppressed, but they are expected to exhibit a harder energy spectrum. Hence, they could dominate the atmospheric neutrino flux at energies above $\sim100\,$TeV. Such a prompt atmospheric flux component has not yet been observed. Therefore, it is an interesting signal in a diffuse neutrino search, but also a background in the search for a diffuse astrophysical neutrino flux. The sensitivity of diffuse neutrino searches with the IceCube Neutrino observatory has reached the level of theoretical expectations of prompt neutrino fluxes, and recent results are presented.
\end{abstract}


\end{frontmatter}



\begin{figure}[t]
  \begin{center}
  \includegraphics[width=0.45\textwidth]{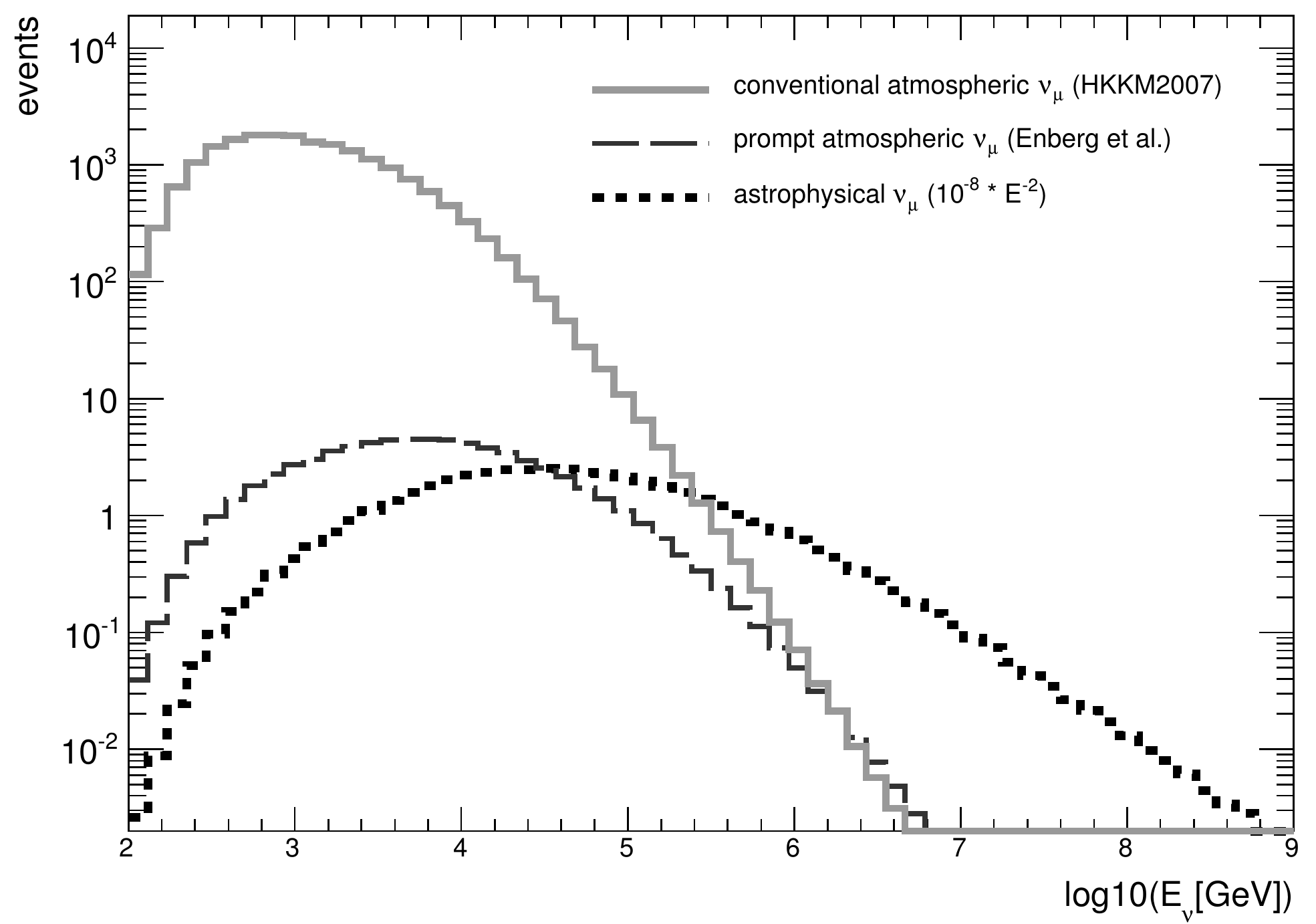}
  \end{center}
  \caption{Distribution of primary neutrino energies for simulated $\nu_{\mu} + \overline \nu_{\mu}$ for conventional atmospheric, prompt atmospheric and astrophysical neutrinos folded with the detection efficiency of this analysis.}
  \label{img:energy}
\end{figure}

Searches for diffuse neutrino fluxes have the goal of identifying astrophysical and prompt atmospheric neutrinos in the measured data. The signatures of a diffuse prompt \cite{Enberg, Martin, Naumov} and extragalactic neutrino signal are their energies. The primary background of conventional atmospheric neutrinos shows an energy spectrum, which is about one power in energy steeper than the spectrum of primary cosmic rays. The reason is the energy dependent competition between the decay and interaction of pions and kaons. At high energies the decay length increases, which makes it more likely that the mesons interact instead of producing muons and neutrinos. Prompt mesons have short lifetimes and decay into neutrinos independent of their energy and arrival direction. Thus, their energy spectrum follows the spectrum of primary cosmic rays. The energy spectrum of astrophysical neutrinos is expected to follow the production spectrum at the cosmic-ray accelerator, which is even harder than the cosmic-ray primary spectrum observed at Earth. Figure \ref{img:energy} illustrates the expected spectrum for prompt and astrophysical neutrinos compared to the conventional atmospheric neutrino background. A diffuse neutrino analysis aims for a precise measurement of the observed neutrino energy distribution in order to search for the characteristic transition to a harder spectrum at very high energies.\vspace{5 pt}

The calculation of the prompt atmospheric flux requires the knowledge of the differential cross sections for $gg \rightarrow c \overline c$ and $q \overline q$-fusion, and therefore the knowledge of parton distribution functions at small values of Bjorken $x$. These are very difficult to measure with collider experiments \cite{Martin}. A measurement of prompt atmospheric muons and neutrinos could therefore complement the knowledge of differential cross sections for parton interactions and structure functions in accelerator physics. This makes the search for a prompt atmospheric neutrino flux an interesting topic.\vspace{5 pt}

The IceCube Neutrino Observatory is a cubic-kilometer neutrino detector installed in the ice at South Pole \cite{IceCube}. It is based on the optical detection of neutrinos, which interact with the ice and produce secondary particles. Such charged particles emit Cherenkov light, which is detected by optical sensors. These sensors are attached to 86 strings, which hold 60 sensors each, with a standard vertical spacing of $17\,$m between sensors and a horizontal distance of $125\,$m between the strings. The data analyzed here were taken between May 2009 and May 2010 with a livetime of $348\,$days, when the IceCube detector was still under construction and in its 59-string configuration. This analysis is based on a selection of muon neutrino events by their characteristic track-like signature. The temporal and spatial hit pattern of light produced along the muon trajectory and recorded by the optical sensors allows a directional reconstruction of the inital neutrino, and the brightness of the event is the basis of an energy reconstruction. The main background consists of atmospheric muons, which outnumber the conventional atmospheric neutrinos by more than five orders of magnitude at trigger level. The muon background is removed, by selecting high-quality upward-going tracks. Such tracks can only originate from neutrinos, which are not absorbed inside the Earth. The final sample consists of $21943$ events with a very small muon background contamination estimated from simulation to be less than $0.2\%$.\vspace{5 pt}

\begin{figure}[t]
  \begin{center}
  \includegraphics[width=0.42\textwidth]{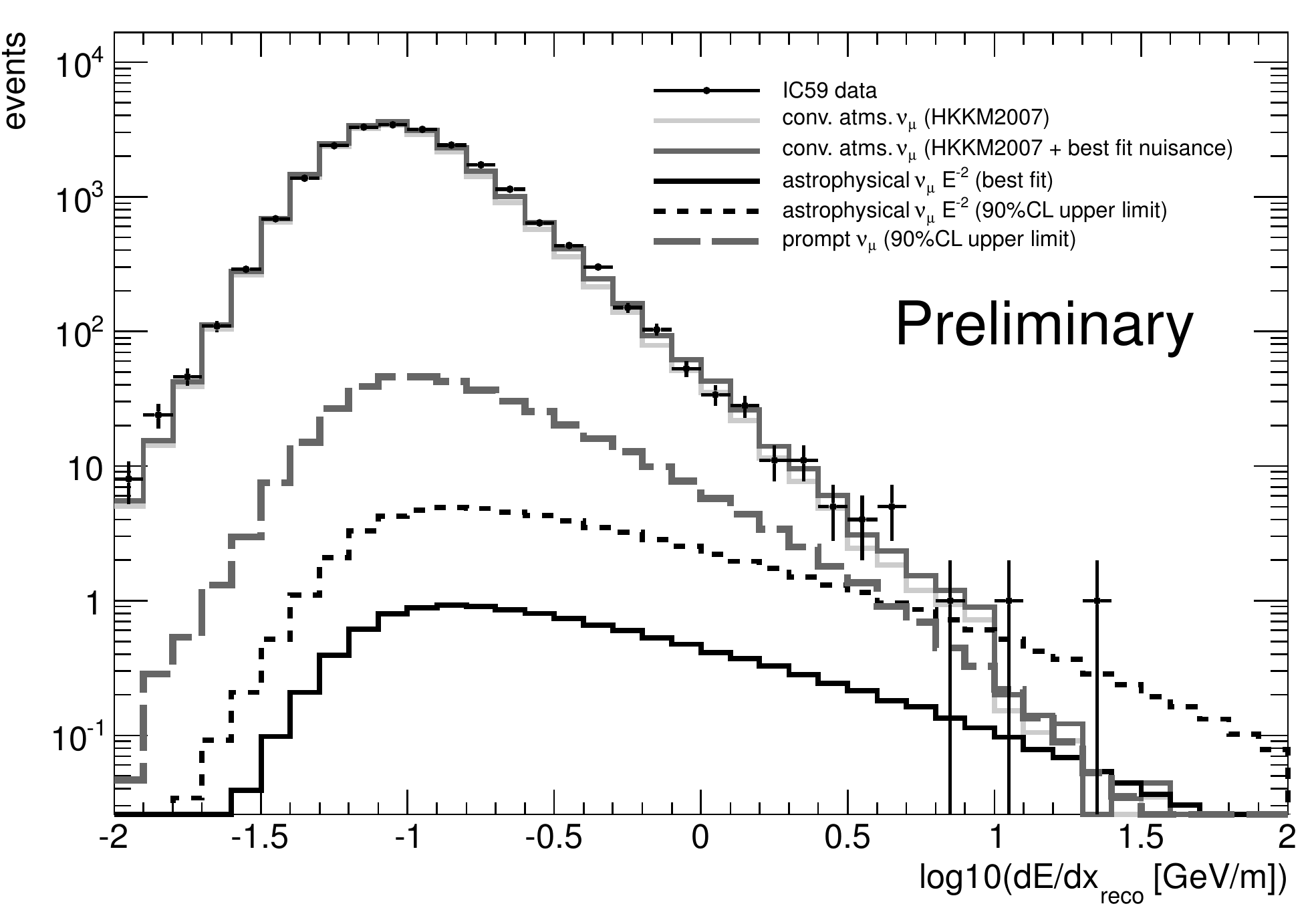}
  \end{center}
  \caption{Reconstructed energy loss of $\nu_{\mu} + \overline \nu_{\mu}$ events measured with the IceCube detector in its 59-string configuration. Additionally shown are the best fit result and upper limit for an extragalactic diffuse neutrino flux and the upper limit on a prompt atmospheric neutrino component.}
  \label{img:dedx}
\end{figure}

Figure \ref{img:dedx} shows the reconstructed muon energy loss distribution measured with IceCube, which is correlated with the initial neutrino energy. The data are analyzed in a two-dimensional likelihood analysis, evaluating the full information from reconstructed neutrino-induced muon energies and arrival directions. The expected zenith angle distribution of prompt atmospheric and astrophysical neutrinos is isotropic, while the expected distribution of arrival directions of conventional atmospheric neutrinos shows characteristic features. This behavior is modified by an energy dependent angular detector acceptance, which increases towards the horizon for high-energy events due to absorption in the Earth. The use of this angular information increases the sensitivity of the analysis. The fit is able to identify contributions from prompt atmospheric and astrophysical neutrinos in the data sample by comparing it with the expected distributions for all three neutrino components. The challenge of this analysis is systematic uncertainties, which distort the expected energy distributions and could hide or fake a signal. Such uncertainties are identified in the modeling of the detector acceptance as well as the theoretical prediction of atmospheric neutrino spectra. These uncertainties have been parameterized and implemented as nuisance parameters into the fitting method, in order to quantify their effects on the analyzed distributions in an unbiased way. A particular improvement in understanding the background expectation of atmospheric neutrinos was the inclusion of the cosmic-ray knee into the neutrino flux calculations \cite{GaisserKnee}.\vspace{5 pt}


The measured energy spectrum in Fig.~\ref{img:dedx} largely agrees with the expectation of conventional atmospheric neutrinos, which is shown for the standard prediction from Honda et al.~\cite{Honda07} and for the spectrum modified by the best fit nuisance parameters. The high-energy tail of the distribution shows a different slope appearing as a small excess of events, which is at the level of $1.8\,\sigma$ compared to the expectation of a pure conventional atmospheric neutrino sample.\vspace{5 pt}

\begin{figure}
  \begin{center}
  \includegraphics[width=0.45\textwidth]{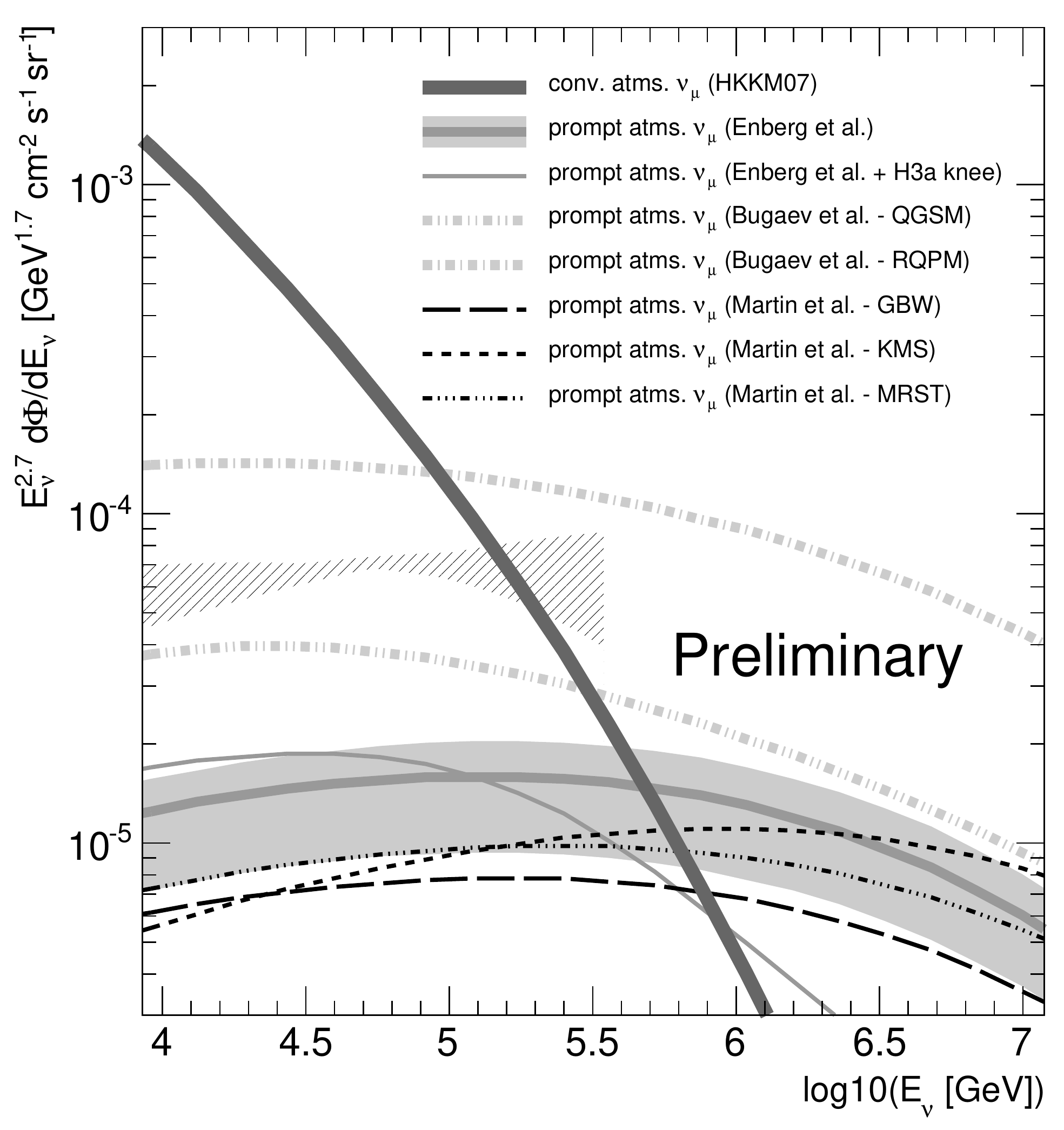}
  \end{center}
  \caption{Predictions for $\nu_{\mu} + \overline \nu_{\mu}$ prompt atmospheric fluxes in comparison to the expected flux of conventional atmospheric neutrinos. The band around the Enberg et al.~prediction marks its theoretical uncertainty. The hatched area represents the envelope containing all limits on the different predictions.}
  \label{img:limitprompt}
\end{figure}

The best fit for prompt atmospheric neutrinos is zero, while the best fit for an extragalactic neutrino flux is non-zero but non-significant. The upper limit at $90\%$ confidence level derived from this analysis on the prompt flux is a factor of $3.8$ larger than the flux calculated by Enberg et al.~\cite{Enberg}, which has been modified for an improved parameterization of the primary cosmic-ray spectrum and composition \cite{GaisserKnee}. This analysis produces individual upper limits valid up to $360\,$TeV, which is the end of the sensitive energy range defined by a worsening of the analysis sensitivity by $5\%$. The limits on each of the models are similar in normalization but follow the slightly different shapes of the models. Figure \ref{img:limitprompt} shows the limits on several prompt neutrino flux predictions in comparison to prompt flux expectations. These limits are below the prediction by Bugaev et al.~(RQPM) \cite{Naumov}, but other prompt neutrino flux predictions are not yet in reach with the current sensitivity.\vspace{5 pt}

\begin{figure}
  \begin{center}
  \includegraphics[width=0.45\textwidth]{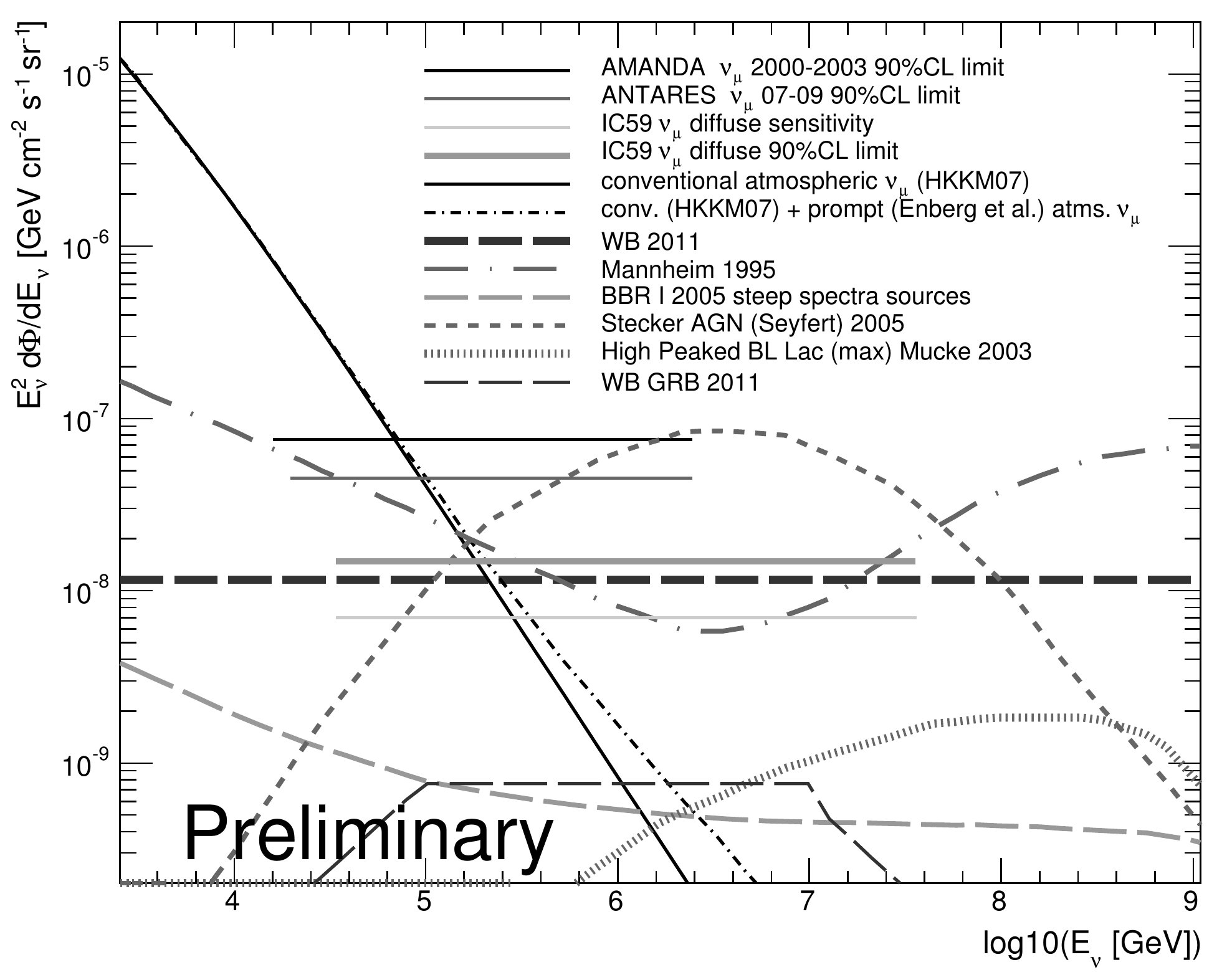}
  \end{center}
  \caption{Limit on an extragalactic $\nu_{\mu} + \overline \nu_{\mu}$ diffuse neutrino flux from this analysis in comparison to limits from previous experiments \cite{Antares, AMANDA} and flux predictions for extragalactic neutrino fluxes for different source classes \cite{Mannheim, Becker, Stecker, Muecke, WBGRB}.}
  \label{img:limitdiffuse}
\end{figure}

The preliminary upper limit derived on a generic astrophysical $E^{-2}$ power law spectrum is $E^2 d\Phi/dE = 1.4\cdot 10^{-8}\,$GeV$\,$cm$^{-2}\,$s$^{-1}\,$sr$^{-1}$, which is slightly above the Waxman-Bahcall bound. The limit on a diffuse astrophysical flux is presented in Fig. \ref{img:limitdiffuse}.\vspace{5 pt}

The completed IceCube detector will provide much higher statistics than this data sample and an expansion to the higher-energy regime. Events with cascade signatures allow a very precise energy reconstruction and smaller background rates from conventional atmospheric muons and neutrinos. In a combined analysis, IceCube will be able to challenge prompt neutrino flux predictions as well as astrophysical flux models within several years of operation.




\nocite{*}
\bibliographystyle{elsarticle-num}
\bibliography{ref}

\begin{thebibliography}{10}
\expandafter\ifx\csname url\endcsname\relax
  \def\url#1{\texttt{#1}}\fi
\expandafter\ifx\csname urlprefix\endcsname\relax\def\urlprefix{URL }\fi
\expandafter\ifx\csname href\endcsname\relax
  \def\href#1#2{#2} \def\path#1{#1}\fi

\bibitem{Enberg}
R.~Enberg, M.~H. Reno, I.~Sarcevic, Phys. Rev. D 78 (2008) 043005.

\bibitem{Martin}
A.~D. Martin, M.~G. Ryskin, A.~M. Stasto, Acta Phys. Polon. B 34 (2003)
  3273--3304.

\bibitem{Naumov}
E.~Bugaev, V.~Naumov, S.~Sinegovsky, E.~Zaslavskaya, Il Nuovo Cimento 12C 1
  (1989) 41.

\bibitem{IceCube}
A.~{Achterberg et al. (IceCube Collaboration)}, Astropart. Phys. 26 (2006)
  155--173.

\bibitem{GaisserKnee}
T.~K. Gaisser, arXiv:1111.6675.

\bibitem{Honda07}
M.~Honda, T.~Kajita, K.~Kasahara, S.~Midorikawa, T.~Sanuki, Phys. Rev. D 75
  (2007) 043006.

\bibitem{Antares}
J.~A. {Anguilar et al. (ANTARES Collaboration)}, Phys. Lett. B 696 (2011)
  16--22.

\bibitem{AMANDA}
A.~{Achterberg et al. (IceCube Collaboration)}, Phys. Rev. D 76 (2007) 042008.

\bibitem{Mannheim}
K.~Mannheim, Astropart. Phys. 3 (1995) 295.

\bibitem{Becker}
J.~K. Becker, P.~L. Biermann, W.~Rhode, Astropart. Phys. 23 (2005) 355--368.

\bibitem{Stecker}
F.~Stecker, Phys. Rev. D 72 (2005) 107301.

\bibitem{Muecke}
A.~Muecke, R.~Protheroe, R.~Engel, J.~Rachen, T.~Stanev, Astropart. Phys. 18
  (2003) 593--613.

\bibitem{WBGRB}
S.~Razzaque, P.~Meszaros, E.~Waxman, Phys. Rev. D 68 (2003) 083001.

\bibitem{WB11}
E.~Waxman, To be published in Astronomy at the Frontiers of Science.

\bibitem{WB98}
E.~Waxman, J.~Bahcall, Phys. Rev. D 59 (1998) 023002.

\end{thebibliography}







\end{document}